# A COMPARISON OF LINK LAYER ATTACKS ON WIRELESS SENSOR NETWORKS

Dr. Shahriar Mohammadi[1] and Hossein Jadidoleslamy[2]

[1] Information Technology Engineering Group, Department of Industrial Engineering, K.N. Tossi University of Technology, Tehran, Iran
Smohammadi40@yahoo.com
[2] Master of Science Student, Department of Information Technology, University of Guilan, Guilan, Iran
Tanha.hossein@gmail.com

***ABSTRACT***

*Wireless sensor networks (WSNs) have many potential applications [1, 5] and unique challenges. They usually consist of hundreds or thousands small sensor nodes such as MICA2, which operate autonomously; conditions such as cost, invisible deployment and many application domains, lead to small size and limited resources sensors [2]. WSNs are susceptible to many types of link layer attacks [1] and most of traditional networks security techniques are unusable on WSNs [2]; due to wireless and shared nature of communication channel, untrusted transmissions, deployment in open environments, unattended nature and limited resources [1]. So, security is a vital requirement for these networks; but we have to design a proper security mechanism that attends to WSN's constraints and requirements. In this paper, we focus on security of WSNs, divide it (the WSNs security) into four categories and will consider them, include: an overview of WSNs, security in WSNs, the threat model on WSNs, a wide variety of WSNs' link layer attacks and a comparison of them. This work enables us to identify the purpose and capabilities of the attackers; also, the goal and effects of the link layer attacks on WSNs are introduced. Also, this paper discusses known approaches of security detection and defensive mechanisms against the link layer attacks; this would enable it security managers to manage the link layer attacks of WSNs more effectively.*

***KEYWORDS***

*Wireless Sensor Network (WSN), Security, Link Layer, Attacks, Detection, Defensive Mechanism*

## 1. INTRODUCTION

Advances in wireless communications have enabled the development of low-cost and low-power wireless sensor networks (WSNs) [1]. WSNs have many potential applications [1, 5] and unique challenges. They usually are heterogeneous systems contain many small devices, called sensor nodes, that monitoring different environments in cooperative; i.e. sensors cooperate to each other and compose their local data to reach a global view of the environment; sensor nodes also can operate autonomously. In WSNs there are two other components, called "aggregation points" and "base stations" [3], which have more powerful resources than normal sensors. Aggregation points collect information from their nearby sensors, integrate them and then forward to the base stations to process gathered data, as shown in figure1. limitations such as cost, invisible deployment and variety application domains, lead to requiring small size and limited resources (like energy, storage and processing) sensors [2]. Also, WSNs are vulnerable to many types of attacks and due to unsafe and unprotected nature of communication channel [4, 9, 22], untrusted and broadcast transmission media, deployment in hostile environments [1, 5], automated nature and limited resources, the most of security techniques of traditional networks are impossible in WSNs; therefore, security is a vital and complex requirement for these networks. It is necessary to design an appropriate security mechanism for these networks [5, 6], which attending to be WSN's constraints. This security mechanism should cover different security dimension of WSNs, include confidentiality, integrity, availability and authenticity. The main purpose of this paper is presenting an overview of different link layer attacks on WSNs and comparing them together. In this paper, we focus on security of WSNs and classify it into four categories, as follows:

- An overview of WSNs,
- Security in WSNs include security goals, security obstacles and security requirements of WSNs,
- The threat model on WSNs,
- A wide variety of WSN's link layer attacks and comparison them to each other, include classification of WSN's link layer attacks based on threat model and compare them to each other based on their goals, results, strategies, detection and defensive mechanisms;

This work makes us enable to identify the purpose and capabilities of the attackers; also, the goal, final result and effects of the attacks on the WSNs. We also state some available approaches of security detection and defensive mechanisms against these attacks to handle them. The rest of this paper is organized as follows: in section 2 is presented an overview of WSNs; while section 3 focused on security in WSNs and presents a diagram about it; section 4 considers the threat model in WSNs; section 5 includes definitions, strategies and effects of link layer attacks on WSNs; in section 6 is considered WSNs' link layer attacks, their goals, effects, possible detection and defensive mechanisms, and extracts their different features, then classifies the link layer attacks based on extracted features and compares them to each other; and finally, in section 7, we present our conclusion.

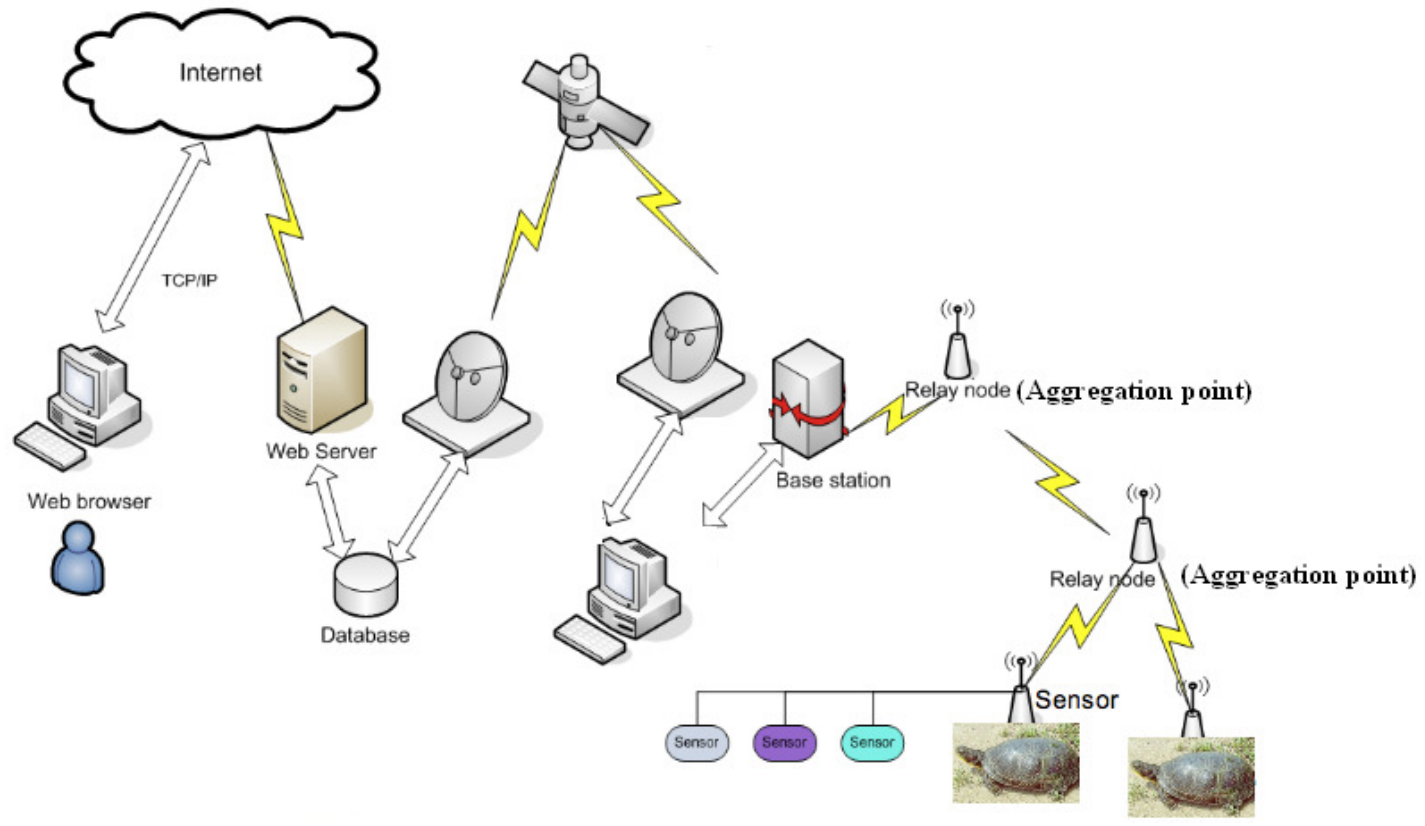

Figure 1. WSN's architecture

## 2. OVERVIEW OF WSNs

In this section, we present an outline of different dimensions of WSNs, such as definition, characteristics, applications, constraints and challenges; as presented in following subsections (subsection 2.1, 2.2, 2.3 and 2.4).

### 2.1. Definition and suppositions of WSNs

A WSN is a heterogeneous system consists of hundreds or thousands low-cost and low-power tiny sensors to monitoring and gathering information from deployment environment in real-time [6, 7, 8]. Common functions of WSNs are including broadcast and multicast, routing, forwarding and route maintenance. The sensor's components are: sensor unit, processing unit, storage/memory unit, power supply unit and wireless radio transceiver; these units are communicating to each other, as shown in following figure (figure2). The existing components on WSN's architecture are including sensor nodes (motes or field devices that are sensing data), network manager, security manager, aggregation points, base stations (access point or gateway) and user/human interface. Besides, there are two approaches in WSN's communication models containing hierarchical WSN versus distributed [6] and homogeneous WSN versus heterogeneous [6]. Some of common suppositions of these networks are:

- Insecure radio links [8, 9, 10],
- Packet injection and replay [8, 9],
- Non tamper resistant [10],
- Many normal sensor nodes (high-density) and low malicious nodes,
- Powerful attackers (laptop-class) [10, 20].

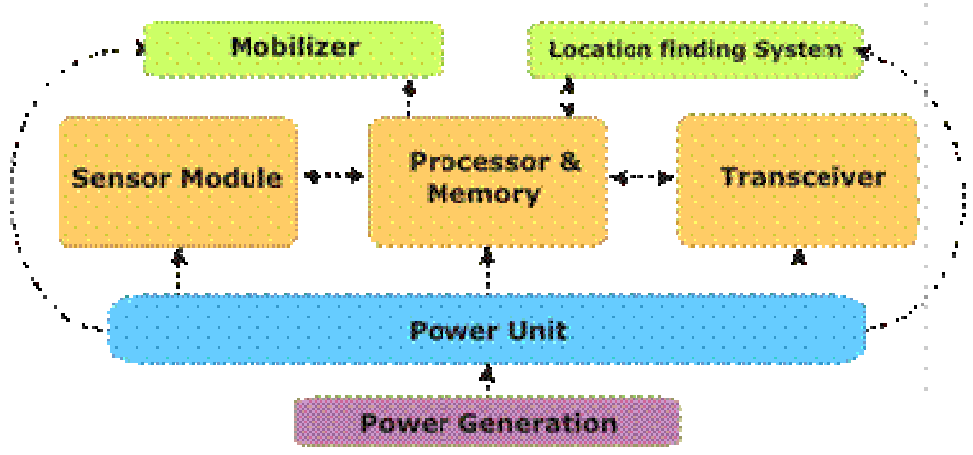

Figure 2. WSN's node architecture

## 2.2. WSNs characteristics and weakness

Most important characteristics of WSNs are including:

- Constant or mobile sensors (mobility),
- Sensor limited resources [4, 18] (limited range radio communication, energy, computational capabilities [4]), **l**ow reliability, wireless communication [4] and immunity,
- Dynamic/unpredictable WSN's topology and self-organization [4, 21],
- Ad-hoc based networks [8, 19] and hop-by-hop communication (multi-hop routing) [11, 12, 21],
- Non-central management, **a**utonomously and **i**nfrastructure-less [8],
- Open/hostile-environment nature [8, 10] and high density;

## 2.3. WSN's applications

In general, there are two kinds of applications for WSNs including, monitoring and tracking [8]; therefore, some of most common applications of these networks are: military, medical, environmental monitoring [2, 6, 8], industrial, infrastructure protection [2, 8], disaster detection and recovery, agriculture, intelligent buildings, law enforcement, transportation and space discovery (as shown in figure3: a and b).

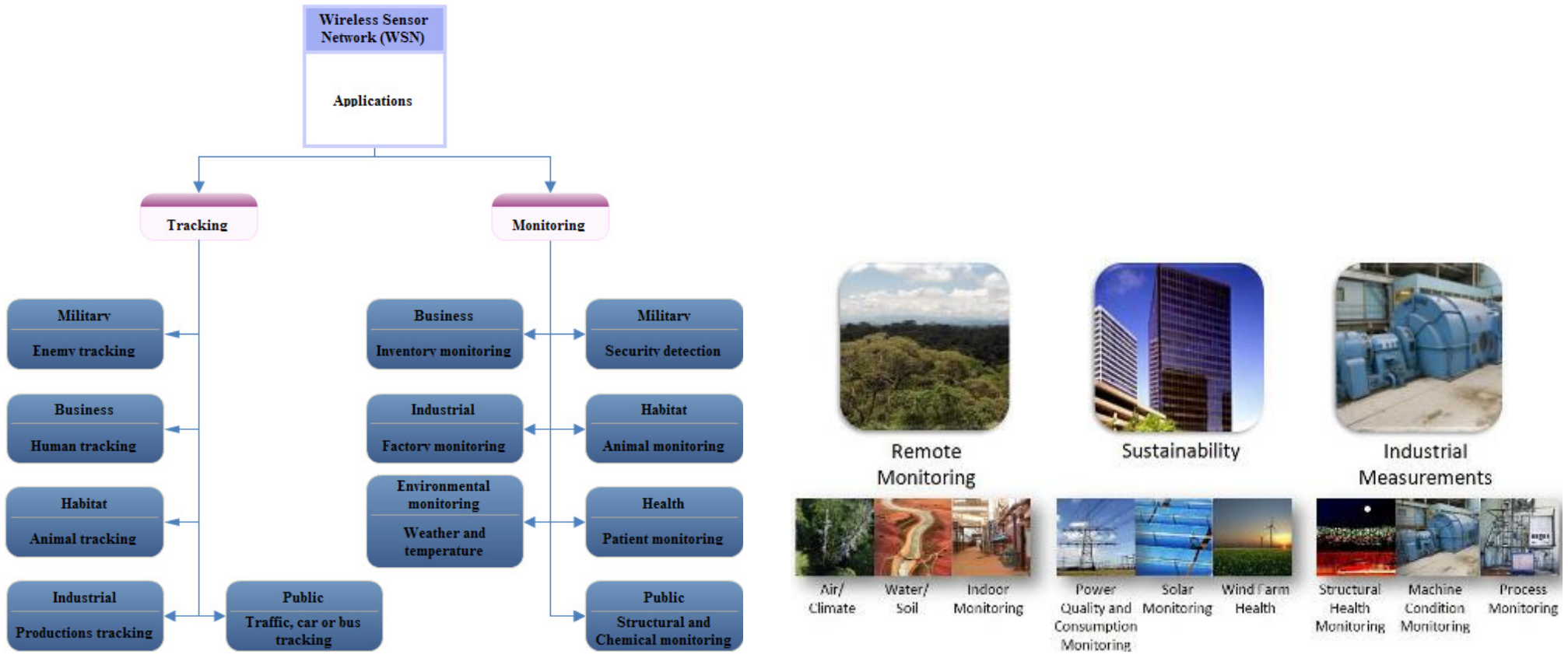

(a) (b)

Figure 3. WSN's applications

## 2.4. Vulnerabilities and challenges of WSNs

WSNs are vulnerable to many kinds of attacks; some of most important reasons are including:

- Theft (reengineering, compromising and replicating),
- Limited capabilities [13, 14] (DoS attacks risks, constraint in using encryption),
- Random deployment (hard pre-configuration) [13, 22],
- Unattended nature [13, 19, 21, 22];

In continue this section states most common challenges and constraints in WSNs; include:

- Deployment on open/dynamic/hostile environments [19, 20, 22] (physical access, capture and node destruction);
- Insider attacks;
- Inapplicable/unusable traditional security techniques [2, 14, 22] (due to limited devices/resources, deploying in open environments and interaction with physical environment);
- Ad-hoc based deployment [19, 20] (dynamic structure and topology, self-organization);
- Resource scarcity/hungry [4, 17, 22] (low and expensive communication/computation/processing resources);
- Immense/large scale (high density, scalable security mechanism requirement);
- Unreliable communication [4, 22] (connectionless packet-based routing ⇨ unreliable transfer, channel broadcast nature ⇨ conflicts, multi-hop routing and network congestion and node processing ⇨ Latency);
- Unattended operation [9, 20] (Exposure of physical attacks, managed remotely, no central management point);
- Redesigning security architectures (distributed and self-organized);
- Increased attacks' risks and vulnerabilities [22], new attacks, increased tiny/embedded devices, multi-hopping routing (selfish) [21];
- Devices with limited capabilities [15, 16], pervasiveness (privacy worries), wireless (medium) [4, 13, 22] and mobility;

## 3. SECURITY IN WSNS

Now, intrusion techniques in WSNs are growth; also there are many methods to disrupt these networks. In WSNs, data accuracy and network health are necessary; because these networks usually use on confidential and sensitive environments. There are three security key points on WSNs, including system (integrity, availability), source (authentication, authorization) and data (integrity, confidentiality). Necessities of security in WSNs are:

- Correctness of network functionality;
- Unusable typical networks protocols [2, 19];
- Limited resources [4, 22, 24];
- Untrusted nodes [4, 19, 20];
- Requiring trusted center for key management [19], to authenticating nodes to each other [25], preventing from existing attacks and selfishness [24, 26] and extending collaboration;

### 3.1. Why security in WSNs?

Security in WSNs is an important, critical issue, necessary and vital requirement, due to:

- WSNs are vulnerable against security attacks [22, 23] (broadcast and wireless nature of transmission medium);
- Nodes deploy on hostile environments [19, 20, 22] (unsafe physically);
- Unattended nature of WSNs [9, 20];

### 3.2. Security issues

This section states the most important discussions on WSNs; it is including key establishment, secrecy, authentication, privacy, robustness to DoS attacks, secure routing and node capture [13, 19];

### 3.3. Security services

There are many security services on WSNs; but some of their common are including encryption and data link layer authentication [17, 19, 20, 24], multi-path routing [19, 21, 24, 25], identity verification, bidirectional link verification [19, 21, 25] and authenticated broadcasts.

### 3.4. Security protocols

This section presents the most common security protocols of WSNs, containing:

- SNEP: Secure network encryption protocol (secure channels for confidentiality, integrity by using authentication, freshness);
- µTESLA [6, 19] (Micro timed, efficient, streaming, loss-tolerant authentication protocol, authentication by using asymmetric authenticated broadcast);
- SPIN (Sensor protocols for information via negotiation): The idea behind SPIN is to name the data using high-level descriptors or meta-data. Before transmission, metadata are exchanged among sensors via a data advertisement mechanism, which is the key feature of SPIN. Each node upon receiving new data, advertises it to its neighbors and interested neighbors, i.e. those who do not have the data, retrieve the data by sending a request message. There is no standard meta-data format and it is assumed to be application specific. There are three messages defined in SPIN to exchange data between nodes, include: ADV message to allow a sensor to advertise a particular meta-data, REQ message to request the specific data and DATA message that carry the actual data [11, 21];
- Broadcasts of end-to-end encrypted packets [24, 25] (authentication, integrity, confidentiality, replay);

As figure4 shows, the most important dimensions of security in WSNs are including security goals, obstacles, constraints, security threats, security mechanisms and security classes; however, this paper considers only star spangled parts/blocks to classify and compare WSNs' link layer attacks based on them; i.e. security threats (including availability, authenticity, integrity and confidentiality) and security classes (containing interruption, interception, modification and fabrication); as shown in table3.

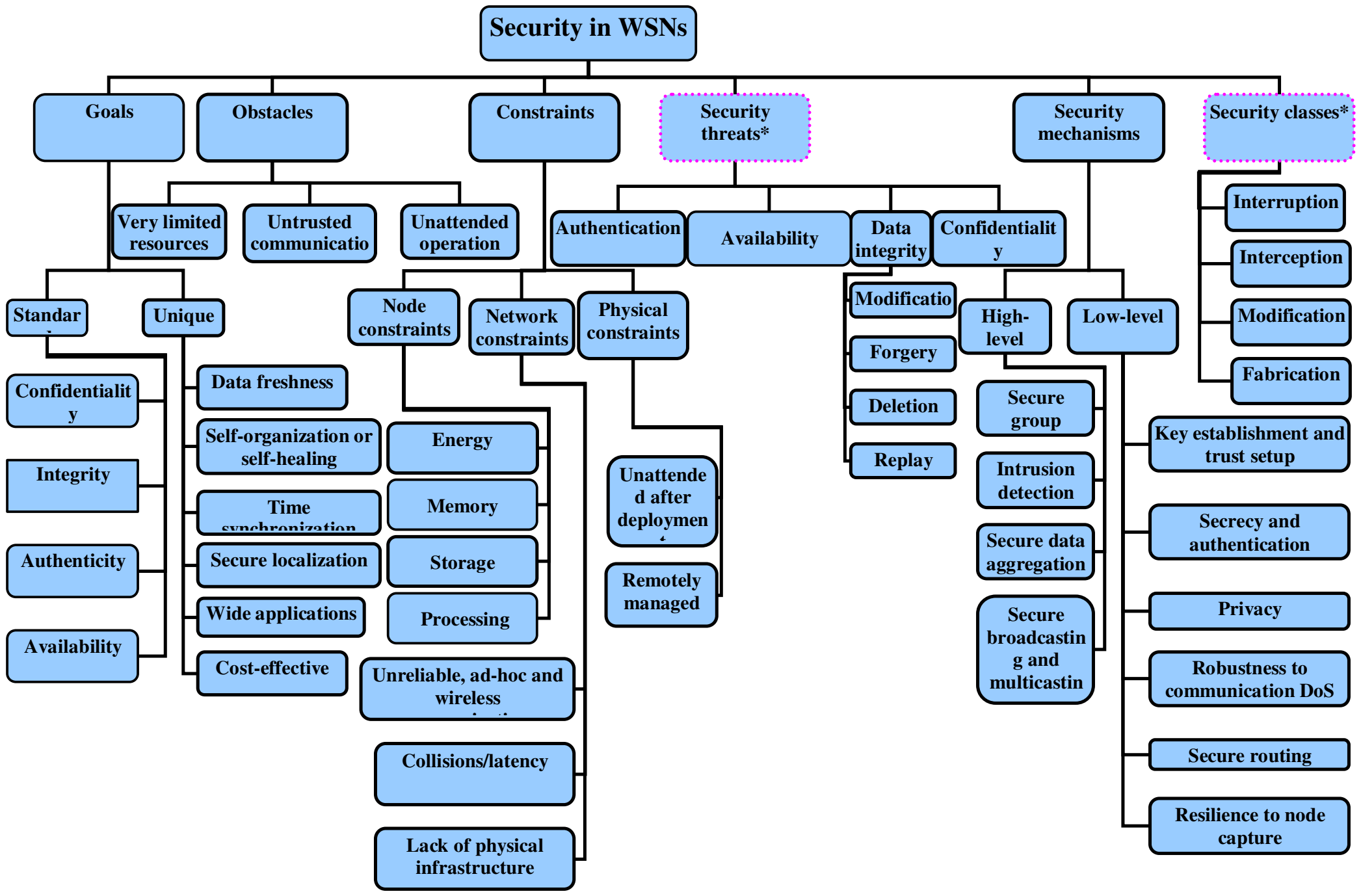

Figure 4. Security in WSNs

## 4. THREAT MODEL IN WSNs

There are many classes of WSNs' attacks based on nature and goals of attacks or attackers; but, in this section we present and compare their most important classes (called threat model of WSNs); as presented in following subsections (subsection 4.1, 4.2, 4.3 and 4.4).

### 4.1. Attacks based on damage/access level

In this subsection is presented the classifications of WSNs' link layer attacks based on their damage level or attacker's access level, including:

**4.1.1. Active attacker:** this kind of attacker does operations, such as:

- Injecting faulty data into the WSN;
- Impersonating [2, 8];
- Packet modification [19];

- Unauthorized access, monitor, eavesdrop and modify resources and data stream;
- Creating hole in security protocols [20];
- Overloading the WSN;

Some of most goals and effects of these attacks are:
- The WSN functionality disruption;
- The WSN performance degradation;
- Sensor nodes destruction;
- Data alteration;
- Inability in use the WSN's services;
- Obstructing the operations or to cut off certain nodes from their neighbors;

**4.1.2. Passive attacker:** passive attacker may do following functions;

- Attacker is similar to a normal node and gathers information from the WSN;
- Monitoring and eavesdropping [2, 20] from communication channel by unauthorized attackers;
- Naturally against privacy;

The goals and effects of this kind of attacker include:
- Eavesdropping, gathering and stealing information;
- Compromised privacy and confidentiality requirements;
- Storing energy by selfish node and to avoid from cooperation;
- The WSN functionality degradation;
- Network partition by non-cooperate in operations;

## 4.2. Attacks based on attacker location

Attacker can be deployed inside or outside the WSN; if the attacker be into the WSN's range, called insider (internal), and if the attacker is deployed out of the WSN's range, called outsider (external). This subsection presented and classified the WSNs' link layer attacks based on attackers' location, including:

**4.2.1. External attacker (outsider):** some of the most common features of this type of attacks are:

- External to the network [2, 19] (from out of the WSN range);
- Device: Mote/Laptop class;
- Committed by illegally parties [2, 7];
- Initiating attacks without even being authenticated;

Some of common effects of these attacks are including:
- Jamming the entire communication of the WSN;
- WSN's resources consumption;
- Triggering DoS attacks;

**4.2.2. Internal attacker (insider):** the meaning of insider attacker is:

- Main challenge in WSNs;
- Sourced from inside of the WSN and access to all other nodes within its range [2, 5, 7];
- Authorized node in the WSN is malicious/compromised;

- Executing malicious data or use of cryptography contents of the legitimate nodes [19, 20];
- Legitimate entity (authenticated) compromising a number of WSN's nodes;

Some of most important goals of these attacks type are:

- Access to cryptography keys or other WSN codes;
- Revealing secret keys;
- A high threat to the functional efficiency of the whole collective;
- Partial/total degradation/disruption;

## 4.3. Attacks based on attacking devices

Attackers can use different types of devices to attack to the WSNs; these devices have different power, radio antenna and other capabilities. There are two common categories of them, including:

**4.3.1. Mote-class attacker:** mote-class attacker is every one that using devices similar to common sensor nodes; this means,

- Occurring from inside the WSN;
- Using WSN's nodes (compromised sensor nodes) or access to similar nodes/motes (which have similar functionality as the WSN's nodes) [7, 8];
- Executing malicious codes/programs;

Mote-class attacker has many goals, such as:

- Jamming radio link;
- Stealing and access to cryptography keys;

**4.3.2. Laptop-class attacker:** laptop-class attacker is every one that using more powerful devices than common sensor nodes, including:

- Main challenge in WSNs;
- Using more powerful devices by attacker, thus access to high bandwidth and low-latency communication channel;
- Traffic injection [2];
- Passive eavesdrop [19] on the entire WSN by a single laptop-class device;
- Replacing legitimate nodes;

Laptop-class attackers have many effects on WSNs, for example:

- Launching more serious attacks and then lead to more serious damage;
- Jamming radio links on the WSN entirely (by using more powerful transmitter);
- Access to high bandwidth and low-latency communication channel;

## 4.4. Attacks based on function (operation)

Link layer attacks in WSNs have been classified into three types, based on their main functionality; this subsection presented them, include:

**4.4.1. Secrecy:** its definition and techniques are:

- Operating stealthy on the communication channel;
- Eavesdropping [4, 20];
- Packet replay, spoofing or modification;
- Injecting false data into the WSN [5, 6];

- Cryptography standard techniques can prevent from these attacks;

Goals and effects of this kind of attacks are:

- Passive eavesdrop;
- Packet replication, spoofing or modification;

**4.4.2. Availability:** this class of attacks known as Denial of Services (DoS) attacks; which leads to WSNs' unavailability, degrade the WSNs' performance or broken it. Some of the most common goals and effects of this attacks' category are including:

- Performance degradation;
- The WSN's services destruction/disruption;
- The WSN useless/unavailable;

**4.4.3. Stealthy:** this kind of attacks is operating stealthy on the communication channel; such as:

- Eavesdropping [2, 8, 20];
- False data injection into the WSN;

The most important effects of these attacks are including:

- Partial/entire degradation/disruption the WSN's services and functionality;

Table 1. Threat model of WSNs

| Attack category/ features | Types | Damage level[1] | Ease of identify[2] | Attacker presence[3] |
|---|---|---|---|---|
| **Based on damage level** | Active attacker | High | Easy | Explicit |
| | Passive attacker | Low | Hard | Implicit |
| **Based on attacker location** | External (outsider) | Low | Medium | Implicit |
| | Internal (insider) | High | Hard | Implicit |
| **Based on attacking devices** | Mote-class attacker | Low | Hard | Implicit |
| | Laptop-class attacker | High | Easy | Explicit |
| **Based on attack function** | Secrecy | High | Hard | Implicit |
| | Availability | High | Hard | Both |
| | Stealthy | High | Hard | Implicit |

As shown in table1, damage level of link layer attacks on WSNs can be high (serious effect on the WSN) or low (limited effect on the WSN); besides, the attackers identification can be easy (possible), medium or hard (impossible), depending on that kind of attack; also the attackers' presence or attacks' effects can be explicit (serious damage) or implicit (for example, eavesdropping).

## 5. DEFINITIONS, STRATEGIES AND EFFECTS OF LINK LAYER ATTACKS ON WSNS

WSNs are designed in layered form; this layered architecture makes these networks susceptible and lead to damage against many kinds of attacks. For each layer, there are some attacks and defensive mechanisms. Thus, WSNs are vulnerable against different link layer attacks, such as DoS attacks, Collision, unfairness and other attacks to link layer protocols [2, 19]; WSNs are susceptible to link layer attacks. Attackers can gain access to transmission media, create radio

[1] damage level: high (serious or more damage than other type) and low (limitary);

[2] ease of identify attackers: easy (possible), medium (depending on attack type) and hard (impossible or not as easy to prevent as other ones);

[3] attacker presence or attack's effect: explicit (more powerful attacker, then more serious damage/harm) and implicit;

interference, prevent from legitimate sensor nodes to communicate/transmit (access to the communication channel) or launch DoS attacks against link layer. Now, in table2 is presented the definitions of link layer attacks on WSNs, and then it classified and compared them to each others based on their strategies and effects.

Table 2. Link layer attacks on WSNs (classification and comparison based on strategies and effects)

| Attack/criteria | Attack definition | Attack techniques | Attack effects |
|---|---|---|---|
| **Node outage** | • Stopping the functionality of WSN's components, such as a sensor node or a cluster-leader; | • Physically[4] ;<br>• Logical[5] ; | • Stop nodes' services;<br>• Take over/compromise the partial/entire the WSN and prevent from some communication;<br>• Impossibility reading gathered information;<br>• Launching other attacks; |
| **Link layer jamming** | • Finding data packet and to jam it[1]; | • Looking at the probability distribution of the inter-arrival times between all types of packets;<br>• This attack can be applied on S-MAC, B-MAC and L-MAC protocols [1]; | • Colliding packets during transmission;<br>• Exhausting nodes' resources;<br>• Confusion; |
| **Collision** | • Message transmission by two nodes on a same frequency [1, 4], simultaneously;<br>• There are 2 types collision: environmental and probabilistic collision; | • Environmental collision;<br>• Probabilistic collision;<br>• Verifying and isolate radio transmissions;<br>• Change packet's fields;<br>• Alter the ack message; | • Interferences [1];<br>• Data/control packets corruption/cripple [1];<br>• Discarding packets;<br>• Energy exhaustion;<br>• Cost effective; |
| **Resource Exhaustion** | • Repeated collisions and continuous retransmission until the sensor node death [1]; | • Continuously retransmission[6];<br>• Interrogation attack (RTS/CTS);<br>• Message modification;<br>• Ack corruption/change[7]; | • Resources exhaustion;<br>• Compromise availability; |
| **Traffic manipulation** | • Regular monitoring transmissions and computing some parameters based on affected MAC protocol carefully ⇨ time adjustment ⇨ transmitting messages just at the moment when normal nodes do so;<br>• Similar to Collision attack; | • Regular monitoring the communication channel and computing require parameters;<br>• Misusing from the wireless nature of communications in WSNs;<br>• Disobeying the coordination rules of MAC schemes in use;<br>• Collision attack techniques;<br>• Unfairness attack techniques;<br>• Continuously collisions and unfairness; | • Excessive packet collisions;<br>• Artificially increased contention;<br>• Decreasing signal quality and network availability;<br>• Aggressively competition for channel usage;<br>• Break the protocols' operations;<br>• Unfair bandwidth usage;<br>• Degradation of the WSN performance;<br>• Traffic distortion;<br>• Effects of collision and unfairness attacks;<br>• Confusion; |

[4] capture and physically damage ⇨ stop functionality;
[5] using other attacks such as collision or exhaustion or unfairness ⇨ node's resources exhaustion ⇨ stop node's functionality;
[6] Continuously retransmit out-of-date/dead/corrupted packets;
[7] Create noise/parasite/interference in acknowledgment messages;

| | | | |
|---|---|---|---|
| **Unfairness** | • Partial DoS attack[8];<br>• Using other attacks such as collision and exhaustion continuously; | • Intermittent application of collision and exhaustion attacks;<br>• Misusing/abusing a cooperative MAC-layer priority mechanism;<br>• Continuously request to access to channel by attacker[9]; | • Decrease utility and efficiency of services;<br>• Nodes' hungry to channel access;<br>• Limiting access to channel and undermine communication channel capacity;<br>• |
| **Acknowledge spoofing** | • An adversary can spoof link layer acknowledgements (ACKs) of overheard packets [10]; | • ACKs replication;<br>• Forging/spoofing link layer ACKs of neighbor nodes; | • False view/information of the WSN;<br>• Launch selective forwarding attack;<br>• Packet loss/corruption; |
| **Sinkhole** | • A special selective forwarding attack;<br>• More complex than blackhole attack;<br>• Attracting [4, 9] or draw the all possible network traffic to a compromised node by placing a malicious node closer to the base station [12] and enabling selective forwarding;<br>• Centralizing traffic into the malicious node [18];<br>• Possible designing another attack during this attack;<br>• Sinkhole detection is very hard[10]; | • Luring [2] or compromising nodes [10];<br>• Tamper with application data along the packet flow path (selective forwarding);<br>• Receiving traffic and altering or fabricating information [12];<br>• Identity spoofing for a short time;<br>• Using the communication pattern;<br>• Creating a large sphere of influence;<br>• Based on used routing protocol: MintRoute or MultiHopLQI protocol; | • Luring and to attract almost all the traffic;<br>• Triggering other attacks, such as eavesdropping, trivial selective forwarding, blackhole and wormhole;<br>• Usurp the base station's position;<br>• Message modification;<br>• Information fabrication and packet dropping;<br>• Suppressed messages in a certain area;<br>• Routing information modification/fake;<br>• Resource exhaustion; |
| **Eavesdropping[11]** | • Detecting the contents of communication by overhearing/stealthy attempt to data; | • Interception;<br>• Abusing of wireless nature of WSNs' transmission medium;<br>• Using powerful resources and strong devices, such as powerful receivers and well designed antennas; | • Launching other attacks (wormhole, blackhole);<br>• Extracting sensitive WSN information;<br>• Delete the privacy protection and reducing data confidentiality; |
| **Impersonation[12]** | • Malicious node impersonates a cluster leader and lures nodes to a wrong position;<br>• Impersonating a node within the path of the data flow of attacker's interest by modifying routing data or implying itself as a trustworthy communication partner to neighboring nodes in parallel; | • The WSN reconfiguration;<br>• Access to encryption keys and authentication information;<br>• Man-in-the-middle attack and fake MAC addresses;<br>• Node replication [23];<br>• Physical access to the WSN;<br>• False or malicious node attack techniques;<br>• Sybil attacks techniques;<br>• Misdirection/misrouting;<br>• Modifying routing information;<br>• Luring/convince nodes; | • Routing information modification;<br>• False sensor readings;<br>• Making network congestion or collapse;<br>• Disclose secret keys;<br>• Network partition;<br>• False and misleading messages generated;<br>• Resources exhaustion;<br>• Degrade the WSN performance;<br>• Invasion;<br>• Carrying out further attacks to disrupt operation of the |

[8] A weaker form of DoS attack;
[9] Cheating/compromising in competition to access to communication channel;
[10] because they use private, invisible and out-of-band channels;
[11] Also called passive information gathering attack; a threat for data confidentiality; the most common attack against privacy; an adversary with powerful resources (powerful receiver and well designed antenna) can gather the data stream from the WSN, if they are not encrypted;
[12] Also called identity spoofing or node replication [23] or multiple identity attacks; identity spoofing and play the role of other one [23]; the attacker assumes the identity of another node in the network, thus receiving messages directed to the node it fakes;

| | | | |
|---|---|---|---|
| | | | WSN;<br>• Confusion and taken over the entire WSN; |
| **Wormholes** | • Tunneling [4, 10] and replicating messages from one location to another through alternative low-latency links [1, 2], that connect two or more points (nodes) of the WSN with fast communication medium [21] (such as Ethernet cable, wireless communication or optical fiber), by colluding two active nodes (laptop-class attackers [2]) in the WSN, by using more powerful communication resources than normal nodes [3, 15] and establishing better real communication channels (tunnel);<br>• Wormhole nodes operate fully invisible [15]; | • Compromising/luring nodes [2] with false and forged routing information;<br>• An attacker locates between two nodes and forwards messages between them;<br>• Using out-of-band or high-bandwidth fast [21] channel;<br>• Wormholes may be used along with Sybil attack;<br>• This attack may combines with selective forwarding or eavesdropping; | • Routing disruption/disorder (false routes, misdirection and forged routing);<br>• False/forged routing information;<br>• Confusion and WSN disruption;<br>• Enable other attacks;<br>• Exploiting the routing race conditions;<br>• Change the network topology;<br>• Prevention of path detection protocol;<br>• Packet destruction/alteration by wormhole nodes;<br>• Changing normal messages stream; |
| **De-synchronization** | • Disrupting the established connections between two legitimate nodes by re-synchronizing their transmission[13]; | • Sending repeatedly forged or false messages;<br>• Re-synchronizing transmissions; | • Disrupt communication;<br>• Go out the synchronization;<br>• Resource exhaustion; |
| **Denial of Service (DoS) attacks** | • A general attack includes several types other attacks in different layers of WSN, simultaneously [28];<br>• Reducing the WSN's availability [19, 28]; | • Physical layer, link layer, routing layer, transport layer and application layer attacks techniques; | • Effects of physical layer, link layer, routing layer, transport layer and application layer attacks; |

## 6. COMPARISON LINK LAYER ATTACKS ON WSNS

WSNs are vulnerable against link layer attacks. Therefore, we have to use some techniques to protect data accuracy, network functionality and its availability. As a result, we require establishing security in WSNs with attention to requirements and limitations of these networks.

### 6.1. Link layer attacks classification based on threat model of WSNs

In this subsection, we have tried to compare the link layer attacks of WSNs based on attacks' nature and effects, attackers' nature and capabilities, and WSN's threat model; as shown in following table (table3).

Table3 shows the most important known attacks on WSNs; this table has three columns, including security class, attack threat and WSNs' threat model. Our purpose of security class is the nature of attacks, includes interruption, interception, modification and fabrication. Attack threat shows which security service attacked or security dimension affected, includes confidentiality, integrity, authenticity and availability. The threat model of WSNs has three sub-columns, that they are presenting attackers' features and capabilities, including based on attacker location (internal/insider or external/outsider), based on attacking devices (mote-class or laptop-class) and based on attacks on WSN's protocols, include active attacks and passive attacks; active attacks are targeting availability (packet drop or resource consumption), integrity (information modification) and authenticity (fabrication); passive attacks are aiming confidentiality (interception).

[13] In link layer: using different neighbors to time synchronization; In transport layer: an established connection between two end points can be disrupted by de-synchronization;

Table 3. WSN's link layer attacks classification based on WSNs' threat model

| Attacks/features | Security class[14] | Attack threat[15] | Threat model[16] | | |
|---|---|---|---|---|---|
| | | | Attacker location | Attacking device | Attacks on WSN's protocols |
| **Node outage** | Modification | Availability, integrity | External | Both | Active |
| **Link layer jamming** | Modification | Availability, integrity | External | Both | Active |
| **Collision** | Modification | Availability, integrity | External | Both | Active |
| **Resource Exhaustion** | Modification | Availability, integrity | External | Both | Active |
| **Traffic manipulation** | Modification | Availability, integrity | External | Both | Active |
| **Unfairness** | Modification | Availability, integrity | External | Both | Active |
| **Acknowledge spoofing** | Fabrication, modification | Integrity, authenticity | Both | Both | Active |
| **Sinkhole** | Modification, fabrication | Availability, integrity, authenticity | Both | Both | Active |
| **Eavesdropping** | Interception | Confidentiality | External | Both | Passive |
| **Impersonation** | Interception, fabrication, modification, | Availability, integrity, confidentiality, authenticity | External | Both | Active |
| **Wormholes** | Fabrication, interception | Confidentiality, authenticity | External | Both | Active |
| **Desynchronization** | Modification, fabrication | Availability, authenticity | External | Both | Active |
| **Denial of Service (DoS) attacks** | Interruption, interception, modification, fabrication | Availability, integrity, confidentiality, authenticity | Both | Both | Active |

Following figure (figure5) shows the nature of WSN's link layer attacks; it compares these attacks based on their nature by presents the percentage of WSNs' link layer attacks which based on interruption, interception, modification or/and fabrication; as a result, the nature of the most of these attacks is modification (almost 85 percent of them).

[14] Security class: the nature of attacks; include interruption, interception, modification and fabrication;
[15] Attack threat: security service attacked; threaten/affected security dimension; include confidentiality, integrity, authenticity and availability;
[16] Threat model: based on attacker location or access level (internal/insider or external/outsider), based on attacking devices (mote-class or laptop-class) and based on damage/attacks on WSN protocols include active attacks (availability (packet drop or resource consumption), integrity (information modification) and authenticity (fabrication)), passive attacks (confidentiality (interception));

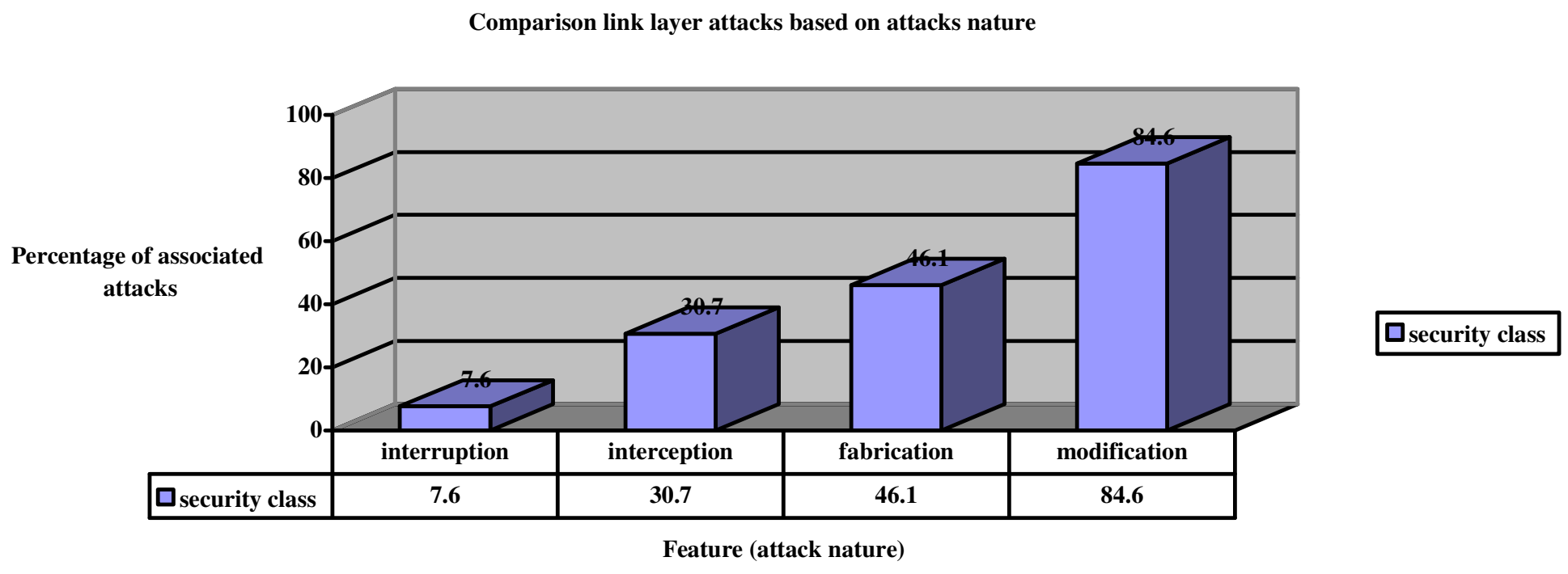

Figure 5. Comparison link layer attacks based on their nature

Following diagram (figure6) shows a comparison of WSNs' link layer attacks based on their security threats factors including confidentiality, integrity, authenticity and availability, in percentage; for example, it presents almost 31 percent of security threat of WSNs' link layer attacks is confidentiality and the nature of 38.4 percent of them is fabrication (fabricating data or identity). As shown in figure6, the aim of the most WSNs' link layer attacks is attacking integrity and availability.

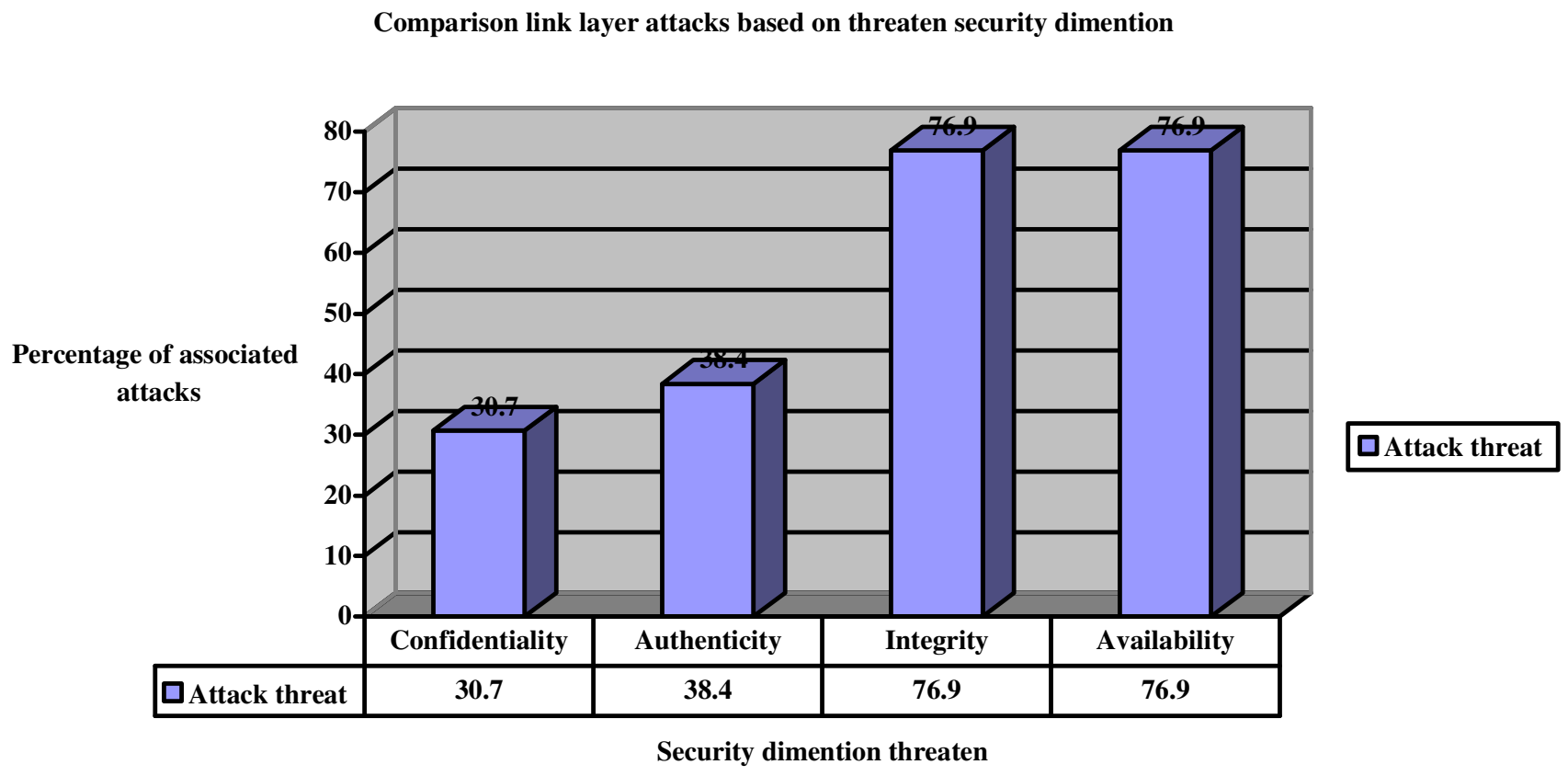

Figure 6. Comparison link layer attacks based on affected security dimension

Following figure (figure7) shows a comparison link layer attacks based on the threat model of WSNs; As shown figure7, the occurred percentage of WSNs' link layer attacks, in attacker location, are 23 percent internal and 100 percent external; i.e. most of WSNs' link layer attacks are occurring from out of WSNs' range and attackers can trigger them by mote-class or laptop-class devices. Also, it presents most of link layer attacks on WSNs are active, except eavesdropping; i.e. almost 92 percent of WSNs' link layer attacks are active. Besides, figure7 shows least attacks on link layer of WSNs are internal attacks.

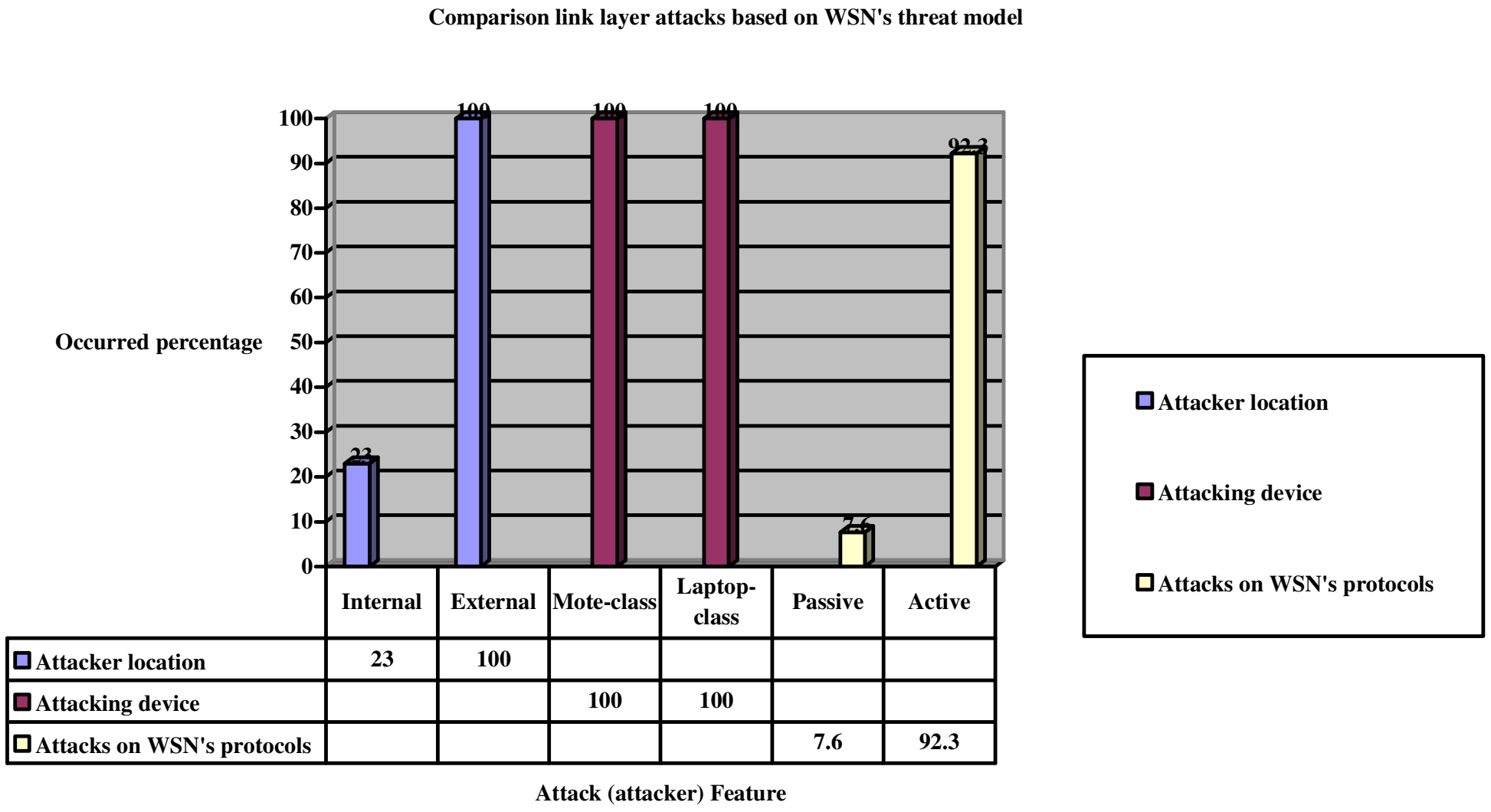

Figure 7. Comparison link layer attacks based on the threat model

## 6.2. Link layer attacks comparison based on their goals and results

In link layer, attacker can disrupt the WSN's functionality by tampering with link layer services such as modifying MAC (Media Access Control) protocol, interference in communication channel and replicating/altering data frames. As shown in table4, it categorizes the link layer attacks of WSNs, based on their goals, effects and results. Also table4 compares WSNs' link layer attacks based on attack or attacker purpose (including passive eavesdrop, disrupt communication, unfairness, authorization and authentication), requirements technical capabilities (such as radio, battery, powerful receiver/antenna and other high-tech and strong attacking devices), vulnerabilities, main target and final result of attacks. Besides, the contributors of all following link layer attacks (shown in table4) are one or many compromised motes, pc or laptop devices on WSNs. The vulnerabilities of these attacks can be physical (hardware), logical or their both; Attacks' main target may be physical (hardware), logical (lis: logical-internal services or lps: logical-provided services) or their both. Final result of these attacks is including passive damage, partial degradation of the WSN functionality and total broken of the WSN's services or functionality.

Table 4. Link layer attacks comparison based on attacks' goals and their results

| Attacks/ features | Purpose[17] | Technical capability | Vulnerability[18] | Main target[19] | Final result[20] |
|---|---|---|---|---|---|
| **Node outage** | Unfairness | - | Logical | lis; lps | PTDB[21] |
| **Link layer jamming [1]** | Disrupt communication | Radio | Logical | lps | PTDB |
| **Collision [1]** | Unfairness | - | Logical | lis; lps | PTDB |
| **Resource Exhaustion [1]** | Unfairness | - | Logical | lis; lps | PTDB |

[17] Purpose: passive eavesdrop, disrupt communication, unfairness, to be authorized, to be authenticated;
[18] Vulnerabilities: physical (hardware), logical;
[19] Main target: physical (hardware), logical (lis: logical-internal services or lps: logical-provided services);
[20] Final result: passive damage, partial degradation of the WSN duty/functionality, service broken/disruption for the entire WSN (partial or total/entire degradation/broken/disruption of the services/resources/functionality of the WSN);
[21] PTDB: Partial/Total Degradation/Broken;

| | | | | | |
|---|---|---|---|---|---|
| **Traffic manipulation** | Unfairness | - | Logical | lis; lps | PTDB |
| **Unfairness** | Unfairness | - | Logical | lis; lps | PTDB |
| **Acknowledge spoofing** | Unfairness | - | Logical | lps | PTDB |
| **Sinkhole [1]** | Unfairness | - | Logical | lps | PTDB |
| **Eavesdropping** | Passive eavesdrop of data | powerful resources and strong devices[22] | Logical | lps | Passive damage; partial degradation |
| **Impersonation** | All purpose | Time and high-tech equipments | Logical; physical | Physical; Logical (lis and lps) | Passive damage; PTDB |
| **Wormholes [1]** | Unfairness; to be authenticated; to be authorized | - | Logical | lps | Passive eavesdrop; PTDB |
| **De-synchronization** | Disrupt communication; unfairness | - | Logical | lis | PTDB |
| **Denial of Service (DoS) attacks** | All purpose | Radio; battery; time and high-tech equipments | Logical; physical | Physical; Logical (lis and lps) | Passive damage; PTDB |

Following figure (figure8) shows that how much percentage of WSNs' link layer attacks are happened by targeting the fairness, confidentiality, authentication, authorization and disrupt communication on WSNs' functionalities, services and resources; for example, almost 85 percent of these attacks are aiming the fairness of WSNs, and then they lead to unfairness.

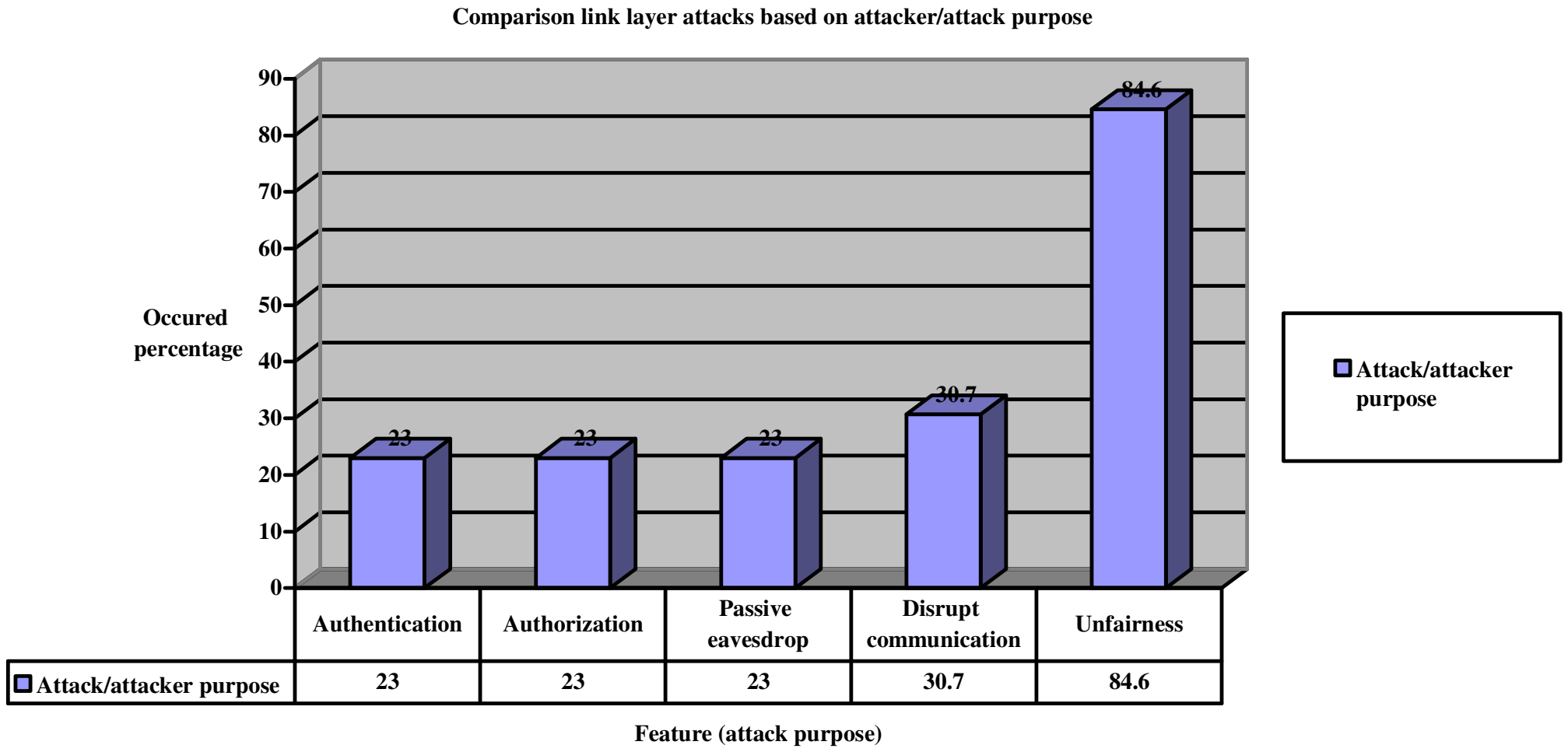

Figure 8. Comparison link layer attacks based on attacks' purpose

Figure9 is presenting the percentage of every one of kinds of link layer attacks vulnerabilities and their main target on WSNs, including: 15.4 percent of them are attacking the WSNs' hardware, 61.5 percent of them are aiming the WSNs' logical-internal services and 92.3 percent are targeting the logical-provided services by WSNs. Thus, most link layer attacks on WSNs have logical vulnerabilities and only almost 15.4 percent of them have physical harm/effects.

[22] such as powerful receiver and well designed antenna;

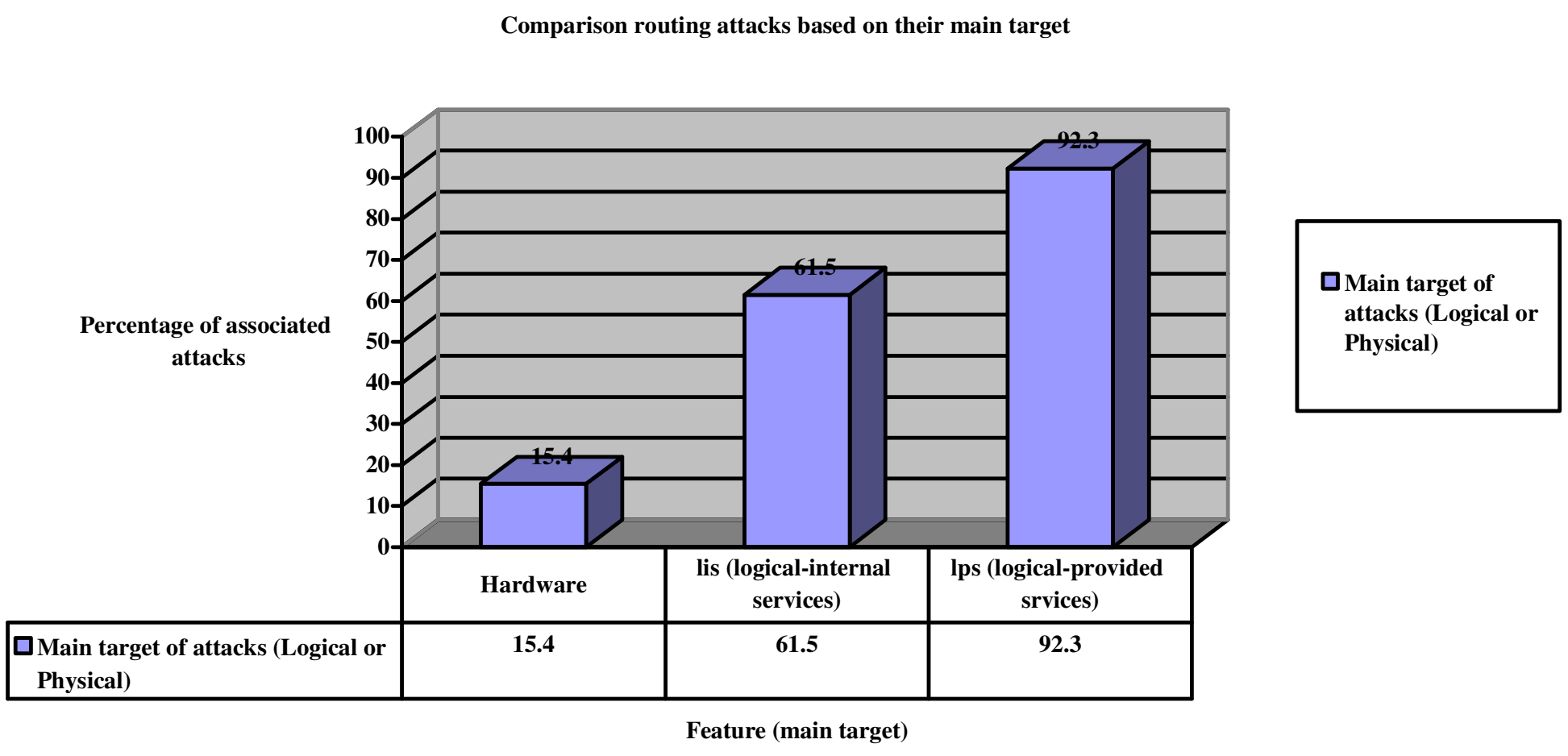

Figure 9. Comparison link layer attacks based on their main target

## 6.3. Detection and defensive strategies of WSNs' link layer attacks

In following table (table5) a classification and comparison of detection and defensive techniques on WSNs' link layer attacks is presented.

Table 5. Link layer attacks on WSNs (classification based on detection and defensive mechanisms)

| Attack/criteria | Detection methods | Defensive mechanisms |
|---|---|---|
| **Node outage** | • Node disconnection from the network;<br>• Regular monitoring and nodes' cooperation;<br>• Existence interference in common operation of node;<br>• Node destruction (physically); | • Providing an alternative path;<br>• Developing appropriate and robust protocols;<br>• Defensive mechanisms against physical and node capture attacks[23]; |
| **Link layer jamming** | • Misbehavior detection techniques[24];<br>• False identity detection techniques; | • Limiting the rate of MAC requests;<br>• Use of small frames;<br>• S-MAC defensive method [1][25];<br>• L-MAC defensive method [1][26];<br>• B-MAC defensive method [1][27];<br>• Identity protection[28];<br>• Link layer encryption; |
| **Collision** | • Misbehavior detection techniques; | • All countermeasures of jamming attack;<br>• Error correction codes (such as CRC codes) [1];<br>• Time diversity; |

[23] Using tamper-proofing/tamper-resistant sensor packages; using special alerting hardware/software to the user; camouflaging/hiding sensors;

[24] include adjustment back-off values, watchdogs/IDS on every node, iterative probing mechanisms, game theory, misbehavior-resilient back-off algorithm, and rating nodes based on replication rate or node's cooperation in communication;

[25] preventing clustering based analysis by narrowing the distance between the two clusters;

[26] making the estimation of the clusters more difficult by changing the slot sizes (used for packet transmission) pseudo-randomly as a function of time;

[27] shortening the preamble in order to make its detection harder;

[28] using cryptography-based authentication or false identity detection techniques such as Radio resource test (Sybil attack), position verification (detecting immobile attackers), code attestation (differing executing code on malicious or compromised node rather than normal nodes ⇨ detecting attackers by validating executing code on nodes), sequence checking and identity association (associating node identity with used keys on communication by that node);

| | | |
|---|---|---|
| **Resource Exhaustion** | • Misbehavior detection techniques; | • Limiting the MAC admission control rate [1];<br>• Random back-offs;<br>• Using Time-Division multiplexing;<br>• limiting the extraneous responses;<br>• Protection of WSN ID and other information; |
| **Traffic manipulation** | • Misbehavior detection techniques; | • Traffic analysis attack defenses;<br>• Collision attack defenses;<br>• Unfairness attack defenses;<br>• Misbehavior detection techniques;<br>• Identity protection;<br>• Link layer encryption;<br>• Limiting the rate of MAC requests;<br>• Use of small frames; |
| **Unfairness** | • Misbehavior detection techniques; | • Use of small frames [1, 2, 4]; |
| **Acknowledge spoofing** | • Misbehavior detection techniques; | • Using another route;<br>• Authentication, link layer encryption and global shared key techniques; |
| **Sinkhole** | • False routing information detection [3, 18];<br>• Cooperating neighboring nodes to each other [18];<br>• Tree structure and verify by tree [18];<br>• Verify by Visual Geographical Map; | • Detection on MintRoute [2];<br>• Geographical routing protocols;<br>• Learning global map (if nodes are static and at known location);<br>• Scalability;<br>• Probabilistic next hop selection;<br>• leveraging global knowledge[29];<br>• Verifying and to trust information that advertised of neighbor nodes;<br>• Authentication, link layer encryption and global shared key techniques;<br>• Routing access restriction (R) [3];<br>• Wormhole detection (W) [3];<br>• Key management (K);<br>• Secure routing (S) [5]; |
| **Eavesdropping** | • Eavesdropping is a passive behavior, thus it is rarely detectable;<br>• Misbehavior detection techniques; | • Access control;<br>• Reduction in sensed data details;<br>• Distributed processing;<br>• Access restriction;<br>• Strong encryption techniques; |
| **Impersonation** | • False identity detection techniques (misbehavior detection techniques);<br>• False routing information detection;<br>• Collision detection techniques; | • Strong and proper authentication techniques;<br>• Using strong data encryption;<br>• Secure routing protocols;<br>• Central certificate authority;<br>• Pair-wise authentication;<br>• Network layer authentication;<br>• Adopt validation techniques;<br>• Identity protection;<br>• Link layer encryption;<br>• Limiting the rate of MAC requests;<br>• Use of small frames for each packet; |
| **Wormholes** | • False routing information detection;<br>• Wormhole detection [15];<br>• Combinational methods [15][30]; | • Packet leach/leashes techniques [1, 21, 27][31];<br>• MAD protocol and OLSR protocol [1, 21]; |

[29] mapping entire network topology by this information and continuously or periodically update the information of base station; misbehavior and serious changes in topology show a compromised node; learning global map (if nodes are static); place nodes at known locations;

[30] such as radio waves and ultrasound, measuring distance between nodes and comparing packet send and receive time with threshold;

[31] Geographical leashes and Temporal leashes ⇨ Physical monitoring of field devices and regular network monitoring by using source routing; monitoring system may use packet leach techniques;

| | | |
|---|---|---|
| | • Packet leashes techniques [21, 27]; | • Directional antennas [1, 26];<br>• Multi-dimensional scaling algorithm (scalability) [1];<br>• Using local neighborhood information [1];<br>• DAWWSEN protocol [2][32];<br>• Designing proper routing protocols (clustering-based and geographical routing protocols);<br>• leveraging global knowledge;<br>• Verifying information that announce of neighbor nodes;<br>• Graphical Position System [26, 27];<br>• Ultrasound [26];<br>• Global clock synchronization[33];<br>• Combinational methods (such as radio waves and ultrasound);<br>• Authentication, link layer encryption and global shared key techniques;<br>• (R), (W), (K), (S) [3, 5]; |
| **De-synchronization** | • Strong and un-forgeable authentication mechanisms; | • Strong authentication mechanisms[34];<br>• Time synchronization, cooperatively[35];<br>• Maintaining proper timing; |
| **Denial of Service (DoS) attacks** | • Detection methods of physical layer, link layer, routing layer, transport layer and application layer attacks; | • Defensive mechanisms of physical layer, link layer, routing layer, transport layer and application layer attacks; |

## 7. CONCLUSION

Security is a vital requirement and complex feature to deploy and extend WSNs in different application domains. The most security link layer attacks are targeting network security dimensions such as integrity, confidentiality, authenticity and availability.

In this paper, we analyze different dimensions of WSN's security, present a wide variety of WSNs' link layer attacks and classify them; our approach to classify and compare the WSN's link layer attacks based on different extracted features of WSN's link layer, attacks' and attackers' properties, such as the threat model of WSNs, link layer attacks' nature, goals and results, their strategies and effects and finally their associated detection and defensive techniques against these attacks to handle them, independently and comprehensively. Table6 presents how much percentage of WSNs' link layer attacks are occurring based on any one attacks' classifications features. Figure10 shows most affected features of WSNs' link layer attacks. Our most important findings are including:

- Discussion typical WSNs' link layer attacks along with their characteristics, in comprehensive;
- Classification and comprehensive comparison of WSNs' link layer attacks to each other;
- Link layer encryption and authentication mechanisms can protect against outsiders, mote-class attackers and link layer attacks such as link layer jamming, traffic manipulation and acknowledgement spoofing;
- Encryption is not enough and inefficient for inside attacks and laptop-class attackers; but clustering protocols can provide most secure solutions against inside attacks and compromised nodes;
- The link layer attacks are often launching combinational;

[32] suspicious node detection by signal strength; a proactive routing protocol based on the hierarchical tree construction;
[33] Using tight clock synchronization, but unfeasible for the majority of WSNs;
[34] to control the identity and the integrity of packets; exchanging packets that are authenticated (including all control fields in the transport protocol header);
[35] Using different neighbors for time synchronization;

- The different kinds of link layer attacks may be used same strategies;
- The same type of defensive mechanisms can be used in multiple link layer attacks, such as misbehavior detection;
- The accuracy of solutions against link layer attacks depends on the characteristics of the WSN's application domain;
- As presented in table6, 84.6 percent of link layer attacks' nature is modification; 30.7 percent of link layer attacks threaten confidentiality, etc;
- As shown in figure10, the nature of 84.6 percent of WSNs' link layer attacks is modification; 76.9 percent of them are targeting integrity and availability; most of these attacks are out of the WSNs' range (external: 100 percent) and lead to high-level damages (active attacks: 92.3 percent); 84.6 percent of attacks' purpose is unfairness; 92.3 percent of link layer attacks' main target is WSNs' logical provided services;

This work makes us enable to identify the purpose and capabilities of the attackers; also the goal, final result and effects of the attacks on the WSNs' functionality. The next step of our work is considering other attacks on WSNs. We hope by reading this paper, readers can have a better view of link layer attacks and aware from some defensive techniques against them; as a result, they can take better and more extensive security mechanisms to design secure WSNs.

Table 6. Occurred percentage of each attacks' classification features

| **Attack or attacker feature** | | **Criteria** | **Percent (percentage of occurred)** |
|---|---|---|---|
| **Security class** | | Interruption | 7.6 |
| | | Interception | 30.7 |
| | | Modification | 84.6 |
| | | Fabrication | 46.1 |
| **Attack threat** | | Confidentiality | 30.7 |
| | | Integrity | 76.9 |
| | | Availability | 76.9 |
| | | Authenticity | 38.4 |
| **Threat model** | **Attacker location** | Internal | 23 |
| | | External | 100 |
| | **Attacking device** | Mote-class | 100 |
| | | Laptop-class | 100 |
| | **Attacks on WSN's protocols** | Passive | 7.6 |
| | | Active | 92.3 |
| **Attacker purpose** | | Disrupt communication | 30.7 |
| | | Authentication | 23 |
| | | Authorization | 23 |
| | | Passive eavesdrop | 23 |
| | | Unfairness | 84.6 |
| **Attack main target** | | Physical (hardware) | 15.4 |
| | | Logical-internal services | 61.5 |
| | | Logical-provided services | 92.3 |

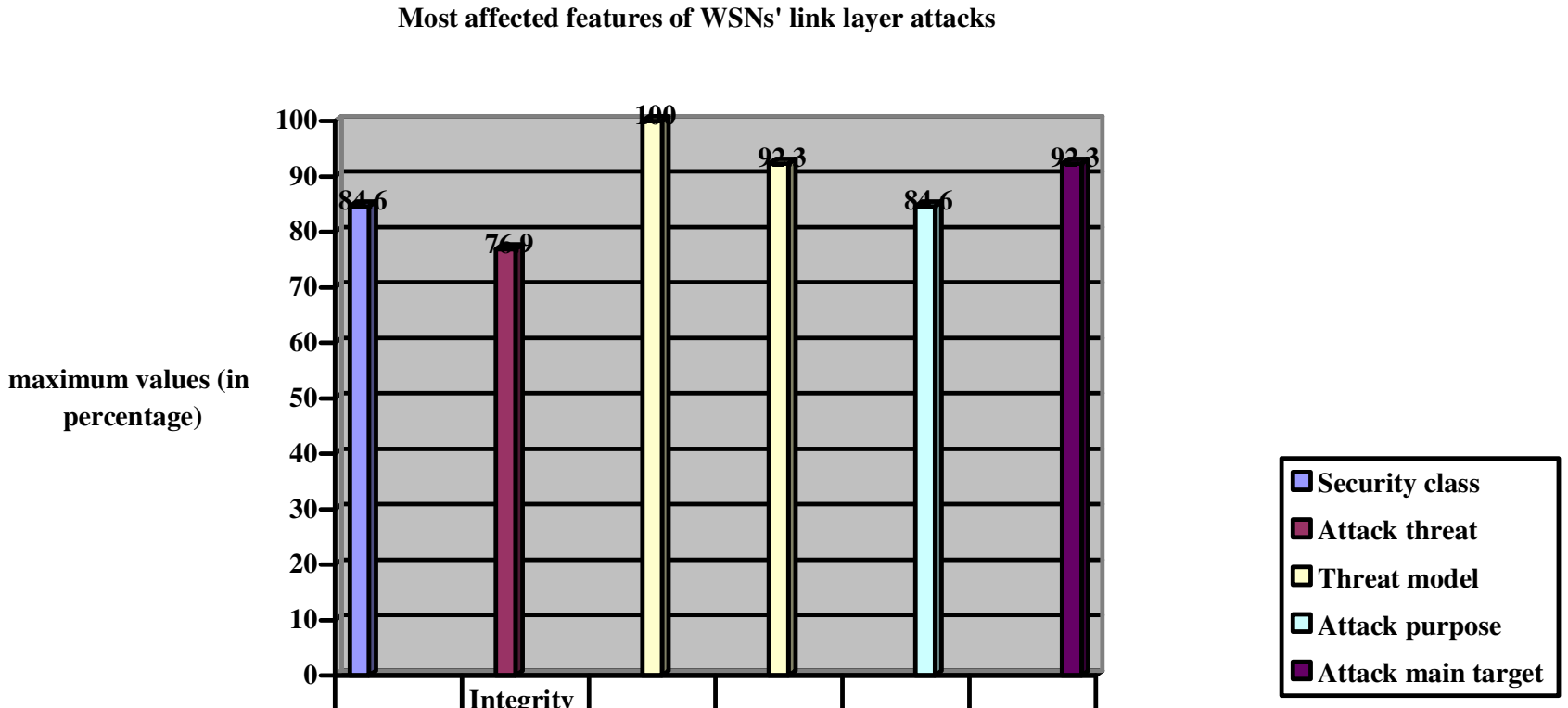

| | Modification | Integrity and Availabili | External | Active | Unfairness | lps |
|---|---|---|---|---|---|---|
| Security class | 84.6 | | | | | |
| Attack threat | | 76.9 | | | | |
| Threat model | | | 100 | 92.3 | | |
| Attack purpose | | | | | 84.6 | |
| Attack main target | | | | | | 92.3 |

most important features of attacks

Figure 10. Most affected features (have maximum values) on WSNs' link layer attacks

## 8. FUTURE WORKS

We also can research about following topics:

- Securing wireless communication links against eavesdropping, collision and DoS attacks;
- Resources limitations techniques;
- Using public key cryptography and digital signature in WSNs (of course with attention to WSN's constraints);
- Countermeasures for combinational link layer attacks;
- Designing proper link layer (MAC[36]) protocols for WSNs;
- Optimizing existing WSNs' MAC protocols;

[36] Media Access Control (MAC)